\documentstyle{article}

\catcode `\@=11

\begin{document}

\def\refname{\small\bf References}

\begin{center}
{\large\bf Quantization of the Classical Maxwell-Nordstr\"om Fields}
\end{center}

\begin{center}
{\bf J. Koci\'nski}
\end{center}

\begin{small}
\begin{center}
Faculty of Physics, Warsaw University of Technology \\ Koszykowa 75, 00-662 Warszawa, Poland
\end{center}
\end{small}

\vspace{7mm}
{\bf\noindent
Abstract} 
\begin{small}
The classical electromagnetic and gravitomagnetic fields in the vacuum, in $(3+2)$ dimensions,
described by the Maxwell-Nordstr\"om equations, are quantized. These equations are
rederived from the field tensor which follows from a five-dimensional form of the Dirac
equation. The electromagnetic field depends on the customary time
$t$, and the hypothetical gravitomagnetic field depends on the second time variable $u$.
The total field energy is identified with the component $T_{44}$ of
the five-dimensional energy-stress tensor of the electromagnetic and 
gravitomagnetic fields. In the ground state,
the electromagnetic field and the gravitomagnetic field energies cancel out.
The quanta of the gravitomagnetic field have spin 1.
\end{small}
\vspace{7mm}

\vspace{7mm}
{\bf\noindent
1. Introduction}
\vspace{7mm}

The Maxwell-Nordstr\"om equations \cite{Nordstrom} refer to pseudo-orthogonal 
five-\\dimensional space. They were rederived \cite{Kocinski1,Kocinski2}, from the iterated 
form of a five-dimensional Dirac equation \cite{Kocinski3,Kocinski4,Kocinski5}.
In the original Nordstr\"om theory \cite{Nordstrom},
the fifth coordinate is a real spatial coordinate. 
When imaginary character is ascribed to the fifth coordinate, 
the Maxwell-Nordstr\"om equations describe electromagnetic phenomena
together with hypothetical gravitomagnetic phenomena \cite{Kocinski2}. In that theory
we are dealing with electromagnetic field and with gravitomagnetic field. 
The electromagnetic waves are periodic in the customary time variable $t$, while the
gravitomagnetic waves are periodic in the second time variable $u$ \cite{Kocinski2}.
\\
\indent
The quantization of the classical Maxwell-Nordstr\"om fields can be accomplished
in the customary way of quantizing the classical electromagnetic field, provided that
the expressions for the total field energy and field momentum of the electromagnetic and
gravitomagnetic fields are known, and that it is accepted
that the respective two types of waves are periodic in two different times.
The field momentum
of these two fields was already determined in \cite{Kocinski2}. The total field energy
can conveniently be determined by applying Sommerfeld's method \cite{Sommerfeld} in
the calculation of the stress-energy tensor in the Minkowski space.
\\
\indent
We will determine
the stress-energy tensor in the two flat spaces $E(3,2)$ and $E(4,1)$. These are connected
with the groups $SO(3,2)$ and $SO(4,1)$, under which the respective Maxwell-Nordstr\"om
equations are covariant \cite{Kocinski1,Kocinski2}.
The total field energy will be identified with the respective $T_{44}$ components
of the stress-energy tensor for those two cases. The result of calculation of the 
stress-energy tensor plays
a pivotal role for the outcome of the quantization procedure, concerning the resultant
ground-state energy of the electromagnetic and gravitomagnetic
fields. It comes out that the signs of the two energy terms which appear in 
the $T_{44}$ component of the stress-energy tensor, the electromagnetic
energy  and the hypothetical gravitomagnetic energy, can be the same or opposite.
These two terms have the same sign in $(4+1)$ dimensions, when the fifth coordinate is real,
and have opposite signs in $(3+2)$ dimensions, when the fifth coordinate is imaginary. 
In the latter case the ground-state energy of the quantized total field vanishes.

\vspace{7mm}
{\bf\noindent
2. The Five-dimensional Stress-Energy Tensor}
\vspace{7mm}

For the method of determining the stress-energy tensor in $(3+2)$ and $(4+1)$ dimensions.
we refer to Sommerfeld \cite{Sommerfeld}. 
All definitions of the relevant quantities and the respective equations can be 
specified in $(3+2)$ or $(4+1)$ dimensions. The formal difference between the two cases
is that the fifth coordinate and the fifth component of the five-potential are respectively
imaginary or real.
We will perform the calculations in the case of $(3+2)$ dimensions,
and next will indicate in which way are the final results modified when $(4+1)-$dimensional
space-time is considered.
\\
\indent
In $(3+2)-$dimensional pseudo-orthogonal space the coordinates of a point 
are: $x_1, x_2, x_3, x_4=ict$, and
$x_5=icu$, where $c$ denotes the speed of light in the vacuum, and $u$ denotes the second
time coordinate. The form of $x_5$ implies that the speed of the gravitomagnetic waves is
assumed to be equal to 
the speed of light $c$ \cite{Kocinski2}. If another speed $c'=\mbox{const}\,c$ were assumed
in the expression for $x_5$, a constant factor
would appear in the formulas, which is irrelevant for the quantization of the fields, and
for the cancellation of the ground-state energies of electromagnetic and gravitomagnetic
fields. 
\\
\indent
We define the five-potential, \cite{Kocinski2},

\begin{equation}
\vec {\cal A}=(A_1, A_2, A_3, A_4, A_5)=\Big(A_x, A_y, A_z, \frac{i}{c}\phi, \frac{i}{c}
\frac{m}{e}\chi\Big)
\label{eq1}
\end{equation}

\noindent
which consists of the three real components $A_x, A_y, A_z$, which are referred to the 
Cartesian system of coordinates $(x_1, x_2, x_3)$, and of two imaginary components,
which are proportional to a  scalar electric potential $\phi$ and a scalar gravitational
potential $\chi$. Under rotations in $(3+2)$ dimensions,
the components of the five-potential $\vec {\cal A}$ transform like the components
of the five-vector $\vec x=(x_1, x_2, x_3, x_4, x_5)$.
\\
\indent 
The respective components of the five-dimensional field tensor have the form \cite{Kocinski2},

\begin{equation}
\partial_j A_k-\partial_k A_j=B_i, \qquad i,j,k=1,2,3
\label{eq2}
\end{equation}

\begin{equation}
\partial_j A_4-\partial_4 A_j=-\frac{i}{c}E_j, \qquad j=1,2,3
\label{eq3}
\end{equation}

\begin{equation}
\partial_j A_5-\partial_5 A_j=-\frac{i}{c}G_j, \qquad j=1,2,3
\label{eq4}
\end{equation}

\begin{equation}
\partial_4 A_5-\partial_5 A_4=-\frac{1}{c}G_0
\label{eq5}
\end{equation}

\noindent
where $B_i$ and $E_j$ are the components of the magnetic induction and of the electric field,
respectively, while $G_j$ is proportional to the hypothetical
gravitational field $G^{\prime}_J$, \cite{Kocinski2}, namely

\begin{equation}
\vec G=\frac{m}{e}\vec G^{\prime}
\label{eq6}
\end{equation}

\noindent
where $m$ denotes the rest mass and $e$ the electric charge of an electron, 
and $G_0$ is proportional to the Brans-Dicke scalar field in the Kaluza-Klein theory
\cite{Overduin} which here is denoted by $G^{\prime}_0$, namely \cite{Kocinski2},
 
\begin{equation}
G_0=\frac{m}{e}G^{\prime}_0
\label{eq7}
\end{equation}

\noindent
To avoid writing the factor $m/e$ in the following formulas, we will use the 
symbols $\vec G$ and $G_0$. 
\\
\indent
The field six-vectors in the Minkowski space \cite{Sommerfeld} are replaced in $(3+2)$
dimensions by the field ten-vectors. The field ten-vector is defined by \cite{Kocinski2}, 

\begin{equation}
F=(c\vec B, -i\vec E, -i\vec G, -G_0)
\label{eq8}
\end{equation}

\noindent
In turn, the excitation ten-vector is defined by

\begin{equation}
f=\sqrt{\frac{\varepsilon_0}{\mu_0}}F=
(\vec H, -ic\vec D, -i\varepsilon_0 c\vec G, -\varepsilon_0 cG_0)
\label{eq9}
\end{equation}

\noindent
where $\vec B$ denotes the magnetic induction, $\vec E$ - the electric field, $\vec G
=\frac{m}{e}\vec G^{\,\prime}$, with $\vec G^{\,\prime}$ denoting the gravitational field, 
and $G_0=\frac{m}{e}G'_0$, with $G'_0$ being the counterpart of the Brans-Dicke scalar; 
$\vec H$ denotes the magnetic field intensity, $\vec D$
is the electric displacement, $\varepsilon_0$ denotes the electric permeability,
and $\mu_0$ the magnetic susceptibility of the vacuum.
\\
\indent
These two ten-vectors have the following matrix form,

\begin{equation}
F=\left[\begin{array}{rrrrr}
0 & cB_z & -cB_y & -iE_x & -iG_x
\\
-cB_z & 0 & cB_x & -iE_y & -iG_y
\\
cB_y & -cB_x & 0 & -iE_z  & -iG_z
\\
iE_x & iE_y & iE_z & 0 & -G_0
\\
iG_x & iG_y & iG_z & G_0 & 0
\end{array}\right]
\label{eq10}
\end{equation}

\noindent
and

\begin{equation}
f=\left[\begin{array}{rrrrr}
0 & H_z & -H_y & -icD_x & -i\varepsilon_0 cG_x
\\
-H_z & 0 & H_x & -icD_y & -i\varepsilon_0 cG_y
\\
H_y & -H_x & 0 & -icD_z & -i\varepsilon_0 cG_z
\\
icD_x & icD_y & icD_z & 0 & -\varepsilon_0 cG_0
\\
i\varepsilon_0 cG_x & i\varepsilon_0 cG_y & i\varepsilon_0 cG_z & \varepsilon_0 cG_0 & 0
\end{array}\right]
\label{eq11}
\end{equation}

\noindent
In the Minkowski subspace, the ten-vectors in Eqs. (\ref{eq8}) and (\ref{eq9}),
and the respective matrices $F$ and $f$ in Eqs. (\ref{eq10}) and (\ref{eq11}), reduce
to those given in \cite{Sommerfeld}.
\\
\indent
We now define the Lagrangian density $\Lambda$ in the form

\begin{equation}
\Lambda =\frac{1}{2c}f\cdot F=\frac{1}{2}(\vec H\cdot\vec B-\vec D\cdot\vec E)-
\frac{1}{2}\varepsilon_0(\vec G^2-G^{2}_0)
\label{eq12}
\end{equation}

\noindent
where the dot $(\cdot)$ denotes the scalar product of the two ten-vectors,
$F$ and $f$ defined in Eqs. (\ref{eq8}) and (\ref{eq9}).
This Lagrangian density is invariant under the transformations of the $SO(3,2)$ group,
owing to the covariance under that group 
of the ten-vectors appearing in the scalar product in Eq. (\ref{eq12}).
\\
\indent
We now observe that to obtain from the above expressions the respective expressions when
the fifth coordinate $x_5$ is real, as it is in the original Nordstr\"om theory
\cite{Nordstrom,Kocinski1}, we have to omit the factor $i$ by the fifth
component of the five-potential in Eq. (\ref{eq1}), and in Eqs. (\ref{eq8}) through
(\ref{eq12}), and to replace $i\vec G$ or $iG_j$ by $\vec G$ or $iG_j$, respectively, and
$G_0$ by $iG_0$. In the Minkowski space,
the Lagrangian in Eq. (\ref{eq12}) reduces to that given in \cite{Sommerfeld}.
\\
\indent
Extending on five dimensions the calculations in \cite{Sommerfeld},
we define the stress-energy tensor $T$ with the components,

\begin{equation}
T_{nm}=-\frac{1}{c}\sum\limits_{r=1}^5 F_{nr}f_{mr}+\delta_{nm}\Lambda
\label{eq13}
\end{equation}

\noindent
where $F_{nr}$ and $f_{mr}$ are the elements of the matrices in Eqs. (\ref{eq10}) and
(\ref{eq11}), respectively. The tensor properties of this quantity follow from the
behaviour of the 
ten-vectors $f$ and $F$ under the $(3+2)-$dimensional rotations. The tensor $T$ is
symmetric since the ten-vector $f$ is proportional to the ten-vector $F$.
\\
\indent
From Eqs.(\ref{eq12}) and (\ref{eq13}), we calculate
the components of the stress-energy tensor $T$ in the form,

\begin{eqnarray}
T_{11}=H_x B_x + D_x E_x-U-\frac{1}{2}\varepsilon_0(\vec G^2-2G^{2}_x-G^{2}_0)
\nonumber\\
T_{12}=H_x B_y + D_y E_x+\varepsilon_0 G_x G_y
\nonumber\\
T_{13}=H_x B_z + D_x E_z +\varepsilon_0 G_x G_z
\nonumber\\
T_{14}=-\frac{i}{c}(\vec E\times\vec H)_x-i\varepsilon_0 G_x G_0
\nonumber\\
T_{15}=-\frac{i}{c}(\vec G\times\vec H)_x-iD_x G_0
\nonumber\\
T_{22}=H_y B_y +D_y E_y -U-\frac{1}{2}\varepsilon_0(\vec G^2-2G^{2}_y-G^{2}_0)
\nonumber\\
T_{23}=H_y B_z +D_z E_y +\varepsilon_0 G_y G_z
\nonumber\\
T_{24}=-\frac{i}{c}(\vec E\times\vec H)_y-i\varepsilon_0 G_y G_0
\nonumber\\
T_{25}=-\frac{i}{c}(\vec G\times\vec H)_y-iD_y G_0
\nonumber\\
T_{33}=H_z B_z +D_z E_z -U-\frac{1}{2}\varepsilon_0(\vec G^2-2G^{2}_z-G^{2}_0)
\nonumber\\
T_{34}=-\frac{i}{c}(\vec E\times\vec H)_z-i\varepsilon_0 G_z G_0
\nonumber\\
T_{35}=-\frac{i}{c}(\vec G\times\vec H)_z-iD_z G_0
\nonumber\\
T_{44}=U-\frac{1}{2}\varepsilon_0(\vec G^2+G^{2}_0)
\nonumber\\
T_{45}=\vec D\cdot\vec G
\nonumber\\
T_{55}=\frac{1}{2}(\vec H\cdot\vec B-\vec D\cdot\vec E)-\frac{1}{2}
\varepsilon_0(G^{2}_0-\vec G^2)
\label{eq14}
\end{eqnarray}

\noindent
where $U$ denotes the electromagnetic field energy, 

\begin{equation}
U=\frac{1}{2}(\vec H\cdot\vec B+\vec D\cdot\vec E)
\label{eq15}
\end{equation}

\noindent
If the speed of the gravitomagnetic waves were assumed to be $ac$, where $a=\mbox{const}$,
the factor $a^2$ would appear in front of $G^2$ and $G^2_0$ in these formulas.
\
\indent
We observe that the trace of the stress-energy tensor,

\begin{equation}
\sum\limits_{r=1}^5 T_{rr}=\frac{1}{2}(\vec H\cdot\vec B-\vec D\cdot\vec E)
-\frac{1}{2}\varepsilon_0(\vec G^2-G^{2}_0)=\Lambda
\label{eq16}
\end{equation}

\noindent
is invariant under the $(3+2)-$dimensional rotations.
\\
\indent
The form of the $T_{44}$ component of the stress-energy tensor is of special interest,
since it determines the total field energy. In Eqs. (\ref{eq14}), the densities of the
electromagnetic field energy and of the gravitomagnetic field energy appearing in $T_{44}$
have opposite signs. Consequently, it will appear that the ground-state energies of the
two fields cancel out.
\\
\indent
For a real fifth coordinate, owing to the above indicated alterations in the formulas,
connected with the passage from $(3+2$ to $(4+1)$ dimensions, namely,
$i\vec G\rightarrow 
\vec G$ and $G_0\rightarrow iG_0$, (the passage from imaginary to real
$x_5-$coordinate), the two energy densities acquire the same sign. Consequently, the 
ground-state energies of the two fields will not cancel out.

\vspace{7mm}
{\bf\noindent
3. The Quantization of Maxwell-Nordstr\"om Fields}
\vspace{7mm}

We consider the stress-energy tensor in $(3+2)$ dimensions in Eq. (\ref{eq14}).
Its $T_{44}$ component is identified with the total energy density $W$ of the two fields,

\begin{equation}
W=\frac{1}{2}(\vec H\cdot\vec B+\vec D\cdot\vec E)-\frac{1}{2}\varepsilon_0
(\vec G^2+G^{2}_0)
\label{eq17}
\end{equation}

\noindent
The density of field momentum connected with the electromagnetic and gravitomagnetic fields
is determined by the Pointing vector \cite{Kocinski2},

\begin{equation}
\vec P=\frac{1}{c^2}(\vec E\times \vec H+\vec G\times \vec H)
\label{eq18}
\end{equation}

\noindent
We assume, as it was done in \cite{Kocinski2}, that the vector potential $\vec A(\vec r,t,u)$
is equal to the sum of two terms, one depending on time $t$ and the other on time $u$, 

\begin{equation}
\vec A(\vec r, t, u)=\vec A_1(\vec r,t)+\vec A_2(\vec r,u)
\label{eq19}
\end{equation}

\noindent
In the vacuum we can assume that

\begin{equation}
A_4=0,  \qquad G'_0=\mbox{const}
\label{eq20}
\end{equation}

\noindent
The last condition has a counterpart in fixing of the scalar field which appears in the
Kaluza-Klein theory, i.e. the Brans-Dicke field, in order to obtain Einstein
equations of general relativity and Maxwell equations (see \cite{Overduin}).
We notice that the conditions $A_4=0$ and $G'_0=\mbox{const}$, imply the loss of
covariance of the five-potential $\vec {\cal A}$ in Eq. (\ref{eq1}) under the group
$SO(3,2)$.
\\
\indent
Assuming that the electric field in Eq. (\ref{eq3}) depends on time $t$ and 
the gravitational field in Eq. (\ref{eq4}) depends on time $u$, we 
obtain \cite{Kocinski2} from those equations and from Eq. (\ref{eq19}) the expressions 

\begin{equation}
\vec E=-\frac{\partial\vec A_1}{\partial t}, \quad \mbox{and} \quad \vec G=
-\frac{\partial\vec A_2}{\partial u},
\label{eq21}
\end{equation}

\noindent
With $\vec H=\mu^{-1}_0 \vec B$, where $\mu_0$ denotes the magnetic susceptibility of
the vacuum, with $\vec D=\varepsilon_0\vec E$ and $\vec B=
{\rm curl}\,\vec A$, we obtain from Eq.(\ref{eq17}) the expression for the total
energy density $W$ in the form

\begin{equation}
W=\frac{1}{2}\Big[\frac{1}{\mu_0}({\rm curl}\,\vec A)^2+\varepsilon_0[
(\frac{\partial \vec A_1}{\partial t})^2-(\frac{\partial \vec A_2}{\partial u})^2]\Big],
\label{eq22}
\end{equation}

\noindent
and from Eq. (\ref{eq18}) the field momentum density $\vec P$ in the form

\begin{equation}
\vec P=-\frac{1}{c^2}\Big[\frac{\partial \vec A_1}{\partial t}+
\frac{\partial \vec A_2}{\partial u}\Big]\times ({\rm curl}\,\vec A)
\label{eq23}
\end{equation}

\noindent
The total field energy and momentum is obtained from these expressions by integrating 
over space. 
\\
\indent
We expand the vector potential in two series of which the first
is connected with the part of the vector potential depending on the time $t$, and the second
with the part of the vector potential depending on the time $u$,

\begin{eqnarray}
\vec A=\vec A_1 (t)+\vec A_2 (u)=\sqrt{\frac{\mu_{0}c^2}{V}}\Big\{\sum
\limits_{\vec k \mu}\,\vec u_{\vec k\mu}\,
\Big(a_{\vec k \mu}e^{i(\vec k\cdot\vec r-\omega t)}+a^{\ast}_{\vec k,\mu}
e^{-i(\vec k\cdot\vec r-\omega t)}\Big)+
\nonumber\\
\sum\limits_{\vec q,\nu}\,\vec v_{\vec q\nu}\,
\Big(b_{\vec q\nu}e^{i(\vec q\cdot\vec r-\omega' u)}+
b^{\ast}_{\vec q\nu}e^{-i(\vec q\cdot\vec r-\omega' u)}\Big\}=\vec A^{\ast}
\label{eq24}
\end{eqnarray}

\noindent
where $\vec k$ and $\vec q$ are the wave vectors of the electromagnetic 
and gravitomagnetic fields, respectively,
where $a_{\vec k,\mu}$ and $b_{\vec q,\nu}$ are linearily independent, complex amplitudes
and where $\ast$ denotes the conjugate complex quantity. According to Eq. (\ref{eq21}),
the unit vectors $\vec u_{\vec k \mu}$
and $\vec v_{\vec q \nu}$, determine the directions of the fields $\vec E$ and $\vec G$,
respectively. The indices $\mu,\nu$ denote the states of transverse polarization.
The condition ${\rm div}\,\vec A=0$, implies that

\begin{eqnarray}
\vec u_{k,\mu}\cdot \vec k=0,\quad \mu=1,2; \qquad \vec v_{q,\nu}
\cdot \vec q=0,\quad \nu=1,2
\nonumber\\
\vec u_{\vec k 1}\cdot\vec u_{\vec k 2}=0,\quad \vec v_{\vec q 1}\cdot\vec v_{\vec k 2}=0
\label{eq25}
\end{eqnarray}

\noindent
The conditions in Eq.(\ref{eq25}) yield for each term under the two sums the expressions
which in turn lead for each term to the conditions

\begin{equation}
\omega=kc, \qquad \omega'=qc
\label{eq26}
\end{equation}

\indent
We now will perform the consequtive quantization steps in the customary way \cite{Bethe,
Flugge}.
In the energy density expression in Eq. (\ref{eq22}), we firstly
consider the term ${\rm curl}\,\vec A$, which is proportional to the following expression,

\begin{equation}
{\rm curl}\,\vec A \sim i\Big[\vec e_{x}(q_{y}u_{z}-q_{z}u_{y})-
\vec e_{y}(q_{x}u_{z}-q_{z}u_{x})+\vec e_{z}(q_{x}u_{y}-q_{y}u_{x})\Big]
\label{eq27}
\end{equation}

\noindent
We also have

\begin{equation}
\vec q\times\vec u_{q,\lambda}=\vec e_{x}(q_{y}u_{z}-q_{z}u_{y})-
\vec e_{y}(q_{x}u_{z}-q_{z}u_{x})+\vec e_{z}(q_{x}u_{y}-q_{y}u_{x})
\label{eq28}
\end{equation}

\noindent
Taking these two expressions into account we obtain

\begin{eqnarray}
{\rm curl}\,\vec A=\sqrt{\frac{\mu_{0}c^2}{V}}\Big\{\sum\limits_{\vec k,\mu}\Big[i(\vec k\times
\vec u_{\vec k,\mu})a_{\vec k,\mu}e^{i(\vec k\cdot\vec r-\omega t)}-
i(\vec k\times\vec u_{\vec k,\mu})a^{\ast}_{\vec k,\mu}e^{-i(\vec k\cdot\vec r-\omega t)}
\Big]+
\nonumber\\
\sum\limits_{\vec q,\nu}\Big[i(\vec q\times\vec v_{\vec q,\nu})b_{\vec q,\nu}
e^{i(\vec q\cdot\vec r-\omega'u)}-i(\vec q\times\vec v_{\vec q,\nu})
b^{\ast}_{\vec q,\nu}e^{-i(\vec q\cdot\vec r-\omega'u)}\Big]\Big\}
\label{eq29}
\end{eqnarray}

\noindent
From $({\rm curl}\, \vec A)^2$, appearing in the energy expression,
we therefore obtain the expressions containing the following terms:

\begin{equation}
-(\vec k\times\vec u_{k,\mu})\cdot (\vec k'\times\vec u_{k',\mu'})=
(\vec k'\cdot \vec k)(\vec u_{k'}\cdot\vec u_{k})-
(\vec u_{k'}\cdot\vec k)(\vec k'\cdot\vec u_{k})
\label{eq30}
\end{equation}

\begin{equation}
-(\vec k\times\vec u_{k,\mu})\cdot (\vec q^{\,\prime}\times\vec v_{q',\nu'})=
(\vec k\cdot\vec q^{\,\prime})(\vec u_{k\mu}\cdot\vec v_{q'\nu})-
(\vec u_{k\mu}\cdot\vec q^{\,\prime})(\vec k\cdot\vec v_{q'\nu'})
\label{eq31}
\end{equation}

\begin{equation}
-(\vec q\times\vec v_{\vec q \nu})\cdot(\vec k'\times\vec u_{\vec k' \mu'})=
(\vec q\cdot\vec k')(\vec v_{\vec q \nu}\cdot\vec u_{\vec k'\mu'})-
(\vec v_{\vec q \nu}\times\vec k')(\vec q\times\vec u_{\vec k'\mu'})
\label{eq32}
\end{equation}

\begin{equation}
-(\vec q\times\vec v_{q\nu})\cdot (\vec q^{\,\prime}\times\vec v_{q'\nu'})=
(\vec q^{\,\prime}\cdot\vec q)(\vec v_{q'\nu'}\cdot\vec v_{q\nu})-
(\vec v_{q'\nu'}\cdot\vec q)(\vec q^{\,\prime}\cdot\vec v_{q\nu})
\label{eq33}
\end{equation}

\noindent
It was shown in \cite{Kocinski2} that plane gravitomagnetic waves are transverse.
For quantization to be possible, the terms in Eqs. (\ref{eq31}) and (\ref{eq32}) have to 
disappear. We consider the two situations:
(1) The electromagnetic wave and the gravitomagnetic wave move in the same direction, and
the vector $\vec E$ of the electromagnetic wave is perpendicular to the vector $\vec G$
of the gravitomagnetic wave. This means that

\begin{equation}
\vec u_{\vec k'}\cdot \vec v_{\vec q}=
\vec u_{\vec k}\cdot \vec v_{\vec q'}=0
\label{eq34}
\end{equation}

\noindent
(2) These two  waves move in mutually perpendicular directions, and the field vectors
$\vec E$ and $\vec G$ are parallel. In both cases the terms in Eqs. (\ref{eq31}) and 
(\ref{eq32}) disappear.
\\
\indent
We next substitute the expansion of the 
vector potential in Eq. (\ref{eq18}) into the terms $\partial\vec A_1/\partial t$ and
$\partial\vec A_2/\partial u$ 
in the energy expression in Eq. (\ref{eq22}). We then obtain for the electromagnetic terms 
the conditions,

\begin{eqnarray}
-\frac{\omega^2}{c^2}-(\vec k'\cdot\vec k)=0 \quad \mbox{when}\quad \vec k'=-\vec k
\nonumber\\
-\frac{\omega^2}{c^2}-(\vec k'\cdot\vec k)=-2\frac{\omega^2}{c^2}\delta_{\mu\mu'}
\quad\mbox{when}\quad \vec k'=\vec k
\label{eq35}
\end{eqnarray}

\noindent
and for the gravitomagnetic field the conditions,

\begin{eqnarray}
\frac{\omega'^{2}}{c^2}-(\vec q^{\,\prime}\cdot\vec q)=2\frac{\omega'^{2}}{c^2}
\delta_{\nu'\nu} \quad \mbox{when}\quad \vec q^{\,\prime}=-\vec q
\nonumber\\
\frac{\omega'^{2}}{c^2}-(\vec q^{\,\prime}\cdot\vec q)=0 \quad\mbox{when}\quad 
\vec q^{\,\prime}=\vec q
\label{eq36}
\end{eqnarray}

\noindent
Consequently, we obtain the total field energy in the form,

\begin{equation}
W=\sum\limits_{\vec k,\mu}\omega^2(a_{\vec k\mu}\,a^{\ast}_{\vec k\mu}+
a^{\ast}_{\vec k\mu}\,a_{\vec k\mu})-\sum\limits_{\vec q,\nu}\omega'^{2}(b_{\vec q\nu}
\,b^{\ast}_{\vec q\nu}+b^{\ast}_{\vec q\nu}\,b_{\vec q\nu})
\label{eq37}
\end{equation}

\noindent
where the first term represents electromagnetic energy and the second term represents
gravitomagnetic energy. These two terms have opposite signs.
For a real fifth coordinate, both terms would have positive sign. This is a direct consequence
of the expression for the component $T_{44}$ of the stress-energy tensor in Eq. (\ref{eq14}).
\\
\indent
The total field momentum is calculated in an analogous way and we obtain,

\begin{equation}
\vec P=\sum\limits_{\vec k,\mu}\omega\vec k (a_{\vec k\mu}\,a^{\ast}_{\vec k\mu}+
a^{\ast}_{\vec k\mu}a_{\vec k\mu})+\sum\limits_{\vec q,\nu}\omega'\vec q (b_{\vec q\nu}
\,b^{\ast}_{\vec q\nu}+b^{\ast}_{\vec q\nu}\,b_{\vec q\nu})
\label{eq38}
\end{equation}

\noindent
The contributions to the total field momentum from both fields have the same sign.
\\
\indent
We now will  quantize the classical electromagnetic and gravitomagnetic 
fields by replacing the amplitudes 
$a_{\vec k\mu}$ and $b_{\vec q\nu}$ and their complex conjugates by the respectives operators
and their Hermitian conjugates, in the form,

\begin{equation}
a_{\vec k\mu}\rightarrow C_{\vec k}\alpha_{\vec k\mu} \quad \mbox{and} \quad
a^{\ast}_{\vec k\mu}\rightarrow C_{\vec k}\alpha^{\dagger}_{\vec k\mu}
\label{eq39}
\end{equation}
\begin{equation}
b_{\vec q\nu}\rightarrow D_{\vec q}\beta_{\vec q\nu} \quad \mbox{and} \quad
b^{\ast}_{\vec q\nu}\rightarrow D_{\vec q}\beta^{\dagger}_{\vec q\nu}
\label{eq40}
\end{equation}

\noindent
with real normalization factors $C_{\vec k}$ and $D_{\vec q}$. We then obtain the 
operators of total field energy ${\cal W}$ and total field momentum ${\cal P}$ in
the form,

\begin{equation}
{\cal W}=\sum\limits_{\vec k,\mu}\omega_{\vec k}^2 C_{\vec k}^2(\alpha_{\vec k\mu}\,
\alpha^{\dagger}_{\vec k\mu}+\alpha^{\dagger}_{\vec k\mu}\,\alpha_{\vec k\mu})-
\sum\limits_{\vec q,\nu}\omega'^{2}_{\vec q}D_{\vec q}^2(\beta_{\vec q\nu}\,
\beta^{\dagger}_{\vec q\nu}
+\beta^{\dagger}_{\vec q\nu}\,\beta_{\vec q\nu})
\label{eq41}
\end{equation}

\noindent
and

\begin{equation}
{\cal P}=\sum\limits_{\vec k,\mu}\omega_{\vec k}\,\vec k \,C_{\vec k}^2
(\alpha_{\vec k\mu}\,\alpha^{\dagger}_{\vec k\mu}+\alpha^{\dagger}_{\vec k\mu}\,
\alpha_{\vec k\mu})+\sum\limits_{\vec q,\nu}\omega'_{\vec q}\,\vec q \,D_{\vec q}^2
(\beta_{\vec q\nu}\,\beta^{\dagger}_{\vec q\nu}+\beta^{\dagger}_{\vec q\nu}\,\beta_{\vec q\nu})
\label{eq42}
\end{equation}

\noindent
We assume the validity of the commutation relations,
\begin{eqnarray}
\alpha_{\vec k\mu}\,\alpha^{\dagger}_{\vec k\mu}-\alpha^{\dagger}_{\vec k\mu}\,
\alpha_{\vec k\mu} =\delta_{\vec k\vec k'}\delta_{\mu\mu'}
\nonumber\\
\beta_{\vec q\nu}\,\beta^{\dagger}_{\vec q\nu}-\beta^{\dagger}_{\vec q\nu}\,
\beta_{\vec q\nu}=\delta_{\vec q\vec q'}\delta_{\nu\nu'}
\label{eq43}
\end{eqnarray}

\noindent
and that all other combinations of these operators commute. We also write,

\begin{equation}
C_{\vec k}=\sqrt{\frac{\hbar}{2\omega_k}} \quad\mbox{and} \quad
D_{\vec q}=\sqrt{\frac{\hbar}{2\omega'_q}}
\label{eq44}
\end{equation}

\noindent
and the operators ${\cal W}$ and ${\cal P}$ then take the form,

\begin{equation}
{\cal W}=\frac{1}{2}\sum\limits_{\vec k,\mu}\hbar\omega_k (\alpha_{\vec k\mu}\,
\alpha^{\dagger}_{\vec k\mu}+\alpha^{\dagger}_{\vec k\mu}\,\alpha_{\vec k\mu})
-\frac{1}{2}\sum\limits_{\vec q,\nu}\hbar\omega'_q(\beta_{\vec q\nu}\,
\beta^{\dagger}_{\vec q\nu}+
\beta^{\dagger}_{\vec q\nu}\,\beta_{\vec q\nu})
\label{eq45}
\end{equation}

\noindent
and
\begin{equation}
{\cal P}=\sum\limits_{\vec k,\mu}\hbar\,\vec k (\alpha_{\vec k\mu}\alpha^{\dagger}_{\vec k\mu}
+\alpha^{\dagger}_{\vec k\mu}\,\alpha_{\vec k\mu})+
\sum\limits_{\vec q,\nu}\hbar\,\vec q (\beta_{\vec q\nu}\,\beta^{\dagger}_{\vec q\nu}+
\beta^{\dagger}_{\vec q\nu}\,\beta_{\vec q\nu})
\label{eq46}
\end{equation}

\noindent
with the eigenvalues

\begin{equation}
W=\sum\limits_{\vec k,\mu}\hbar\omega (N_{\vec k,\mu}+\frac{1}{2})-
\sum\limits_{\vec q,\nu}\hbar\omega' (N_{\vec q,\nu}+\frac{1}{2})
\label{eq47}
\end{equation}

\noindent
and

\begin{equation}
P=\sum\limits_{\vec k,\mu}\hbar\vec k (N_{\vec k,\mu}+\frac{1}{2})+
\sum\limits_{\vec q,\nu}\hbar\vec q (N_{\vec q,\nu}+\frac{1}{2})
\label{eq48}
\end{equation}

\noindent
We are dealing with quanta of electromagnetic field and with quanta of the
hypothetical gravitomagnetic field.
\\
\indent
From Eq. (\ref{eq47}) it follows that the ground state energy of the system of
harmonic oscillators representing electromagnetic and gravitomagnetic field
is equal to zero. 

\vspace{7mm}
{\bf\noindent
4. The Spin of the Quanta of a Gravitomagnetic Field}
\vspace{7mm}

To determine the spin of a quantum of a gravitomagnetic wave
we follow the argument in \cite{Schweber}, adapted for the present case.
In the vacuum we have, \cite{Kocinski2},

\begin{eqnarray}
{\rm div}\,\vec B(\vec r,u)=0,
\nonumber\\
{\rm div}\,\vec G(\vec r,u)=0,
\nonumber\\
{\rm curl}\,G(\vec r,u)=-\frac{\partial \vec B(\vec r,u)}{\partial u}
\label{eq49}
\end{eqnarray}

\noindent
We define the vector
\begin{equation}
\vec F=\frac{1}{\sqrt{2}}(\vec G+ic\vec B)
\label{eq50}
\end{equation}

\noindent
From Eqs. (\ref{eq49}) and (\ref{eq50}),  we then find that

\begin{eqnarray}
{\rm div}\,\vec F=0,
\nonumber\\
i\partial_{u}\vec F=c\,{\rm curl}\,\vec F
\label{eq51}
\end{eqnarray}

\noindent
With $\varepsilon_{lmn}$ denoting the Levi-Civita tensor, we rewrite the last equation in
the form

\begin{equation}
i\partial_{u}F_{l}=c\varepsilon_{lmn}\partial_{m}F_{n}
\label{eq52}
\end{equation}

\noindent
Introducing the angular momentum matrices,
\begin{eqnarray}
s_1=i\left(
\begin{array}{rrr}
0&0&0\\
0&0&1\\
0&-1&0
\end{array}
\right),
\quad
s_2=i\left(
\begin{array}{rrr}
0&0&-1\\
0&0&0\\
1&0&0
\end{array}
\right)
\nonumber\\
s_3=i\left(
\begin{array}{rrr}
0&1&0\\
-1&0&0\\
0&0&0
\end{array}
\right)
\label{eq53}
\end{eqnarray}

\noindent
with $\partial_{m}=\frac{i}{\hbar}p_m$, we can rewrite Eq.(\ref{eq52}) in the form

\begin{equation}
-c(s_m)_{ln}p_mF_n=i\hbar\partial_{u}F_l
\label{eq54}
\end{equation}

\noindent
or in the form
\begin{equation}
-c(\vec s\cdot\vec p)_{ln}F_n=i\hbar\partial_{u}F_l
\label{eq55}
\end{equation}

\noindent
which can be interpreted as the Schr\"odinger equation for the quantum of
the gravitomagnetic field, with the Hamiltonian,

\begin{equation}
-c(\vec s\cdot\vec p)={\cal H}
\label{eq56}
\end{equation}

\noindent
while $\vec F$ in Eq.(\ref{eq50}) is the wave function.

\indent
The condition ${\rm div}\,\vec F=0$ can be rewritten in the form

\begin{equation}
\vec p\cdot\vec F=0
\label{eq57}
\end{equation}

\noindent
We now verify that the Schr\"odinger equation in Eq.(\ref{eq55}) has a plane wave solution

\begin{equation}
\vec F(\vec r,u)=\vec f_{\vec q}e^{i\vec q\cdot\vec r-\omega'u)}
\label{eq58}
\end{equation}

\noindent
where $\vec f_{\vec q}$ is represented by the column matrix
\begin{equation}
\vec f_{\vec q}=
\left(
\begin{array}{r}
f_x\\
f_y\\
f_z
\end{array}
\right)
\label{eq59}
\end{equation}

\noindent
Introducing into Eq.(\ref{eq55}) the matrices $s_l$ in Eq. (\ref{eq53}) we obtain
the equations

\begin{eqnarray}
\omega'f_x+icq_{z}f_y-icq_{y}f_z=0
\nonumber\\
-icq_{z}f_x+\omega'f_y+icq_{y}f_z=0
\nonumber\\
icq_{y}f_x-icq_{x}f_y+\omega'f_z=0
\label{eq60}
\end{eqnarray}

\noindent
which have nonzero solutions when
\begin{equation}
(\omega')^2=c^2\,q^2
\label{eq61}
\end{equation}

\noindent
There follows the conclusion that we are dealing with the spin-one quantum, of
the gravitomagnetic field.

\vspace{7mm}
{\bf\noindent
5. Conclusions}
\vspace{7mm}

The notion of a second time variable is not alien to physics. A brief review of papers
in which the second time variable is considered was given in \cite{Kocinski2}.
In a Kaluza-Klein theory with two times, the Schwarzschild type solution of the five-dimensional
Einstein equations in the vacuum was determined in \cite{Kocinski6,Kocinski7}.
The two independent parameters of that solution are related with mass and electric charge,
respectively. That solution exhibits a Schwarzschild radius, whose magnitude is predominantly
determined by the electric charge. It was shown that the perihelic motion of a test particle
in four-dimensional relativity has a counterpart in five dimensions in the perinucleic
motion of a negatively-charged test particle. With the quantization conditions of the 
older quantum theory included into the five-dimensional geometry, the perinucleic motion
of an electron leads to the fine structure of line spectra which is analogous to that determined 
by Sommerfeld's formula for hydrogen-like atoms.
\\
\indent
The Maxwell-Nordstr\"om equations were rederived in \cite{Kocinski1,Kocinski2} from the
field tensor
implied by the iterated form of a five-dimensional Dirac equation \cite{Kocinski3,Kocinski4,
Kocinski5}.
From the Maxwell-Nordstr\"om equations with two times there follows the existence of
hypothetical gravitomagnetic phenomena, together with the electromagnetic phenomena. 
In particular, plane gravitomagnetic waves with a transverse polarization are implied by those
equations \cite{Kocinski2}.
\\
\indent
In this paper the quantization of the classical electromagnetic and gravitomagnetic
fields has lead to the conclusion that in $(3+2)$ dimensions, the ground-state energies
of electromagnetic and gravitomagnetic fields in the vacuum cancel out.
It has been shown that the quanta of the hypothetical gravitomagnetic field have spin 1.
\\
\indent
The idea that a remedy for the infinite energy of the ground state of the quantized
electromagnetic field
may be sought in finding another energy which will cancel that infinity out, was expressed
by Wesson \cite{Wesson}. It seems that we are dealing with such a case.

\end{document}